# Generalized Criterion of Maximally Multi-Qubit Entanglement

Xinwei Zha,[1,][*] Chenzhi Yuan,[2] and Yanpeng Zhang[2,][†]

[1]*School of Science, Xi'an University of Posts and Telecommunications, Xi'an 710121, P R China*

[2]*Key Laboratory for Physical Electronics and Devices of the Ministry of Education & Shaanxi Key Lab of Information Photonic Technique, Xi'an Jiaotong University, Xi'an 710049, P R China*

[*] zhxw@xupt.edu.cn; [†] ypzhang@mail.xjtu.edu.cn

We first present a generalized criterion for maximally entangled states of 2, 3, 4，5，6, 8 and in theory to arbitrary-number qubits. By this criterion, some known highly entangled multi-qubit states are examined and a new genuine eight-qubit maximally entangle state is obtained. For the 4, 7 and 8 qubits system in which no maximally multi-qubit entangled states (MMES) is thought to exist before, we find that the proved and most suspected MMESes, are not completely mixed in subsystem with a critical-number qubits, below which the subsystems are all completely mixed. We believe that the new criterion and MMES can play important role in quantum information technology, such as the teleportation and dense coding.

Entanglement is considered to be the central resource for quantum information and computation [1, 2], and numerous theoretical and experimental detailed works have been done in this field [3-6]. As an important feature in the quantum many-body system, the multi-particle entanglement has been intensively investigated since it is significantly different from the trivial extension of biparticle entanglement [7-10], and the characterization and classification of such entanglement now become a issue full of interest [11, 12]. A common feature of entanglement for multipartite quantum system is that its non-local properties do not change under local transformations, so local unitary (LU) transformations invariants has been used for the classification of the multi-qubit states [13, 14].

As perfect entanglement resource in quantum communication and computing, the maximally multi-qubit entangled state (MMES) have received much theoretical attentions. A definition of MMES has been proposed recently, yet the numerical searching method base on this definition can lead to the inexistence of 4 qubits MMES and becomes frustration in finding the 8-qubit MMES [15]. Though there are several known examples of maximally entangled quantum states of 2, 3, 5 and 6 qubits [3, 16, 17, 18]; however, the mathematical structure for multi-qubit entanglement of more than 7 qubits is less clear. In this letter, we first propose a general criterion to determine whether a state is maximally entangled, which operates well on several known highly entangled state and in theory can engage in entanglement in arbitrary $n$-qubit system. We also first discover a genuine 8-qubits entangled state proved to be maximally entangled by this new criterion. In the tough system (4, 7, 8 qubits system) in which no MMES is thought to exist before, the proved (for 4 and 8 qubits) and most suspected MMESes (for 7 qubits) discovered under our new criterion is not completely mixed in a critical number qubit (2, 3 and 4 for 4, 7 and 8, respectively) subsystem, but is completely mixed in the subsystem with qubits less than the critical number.

At first, we start from the definition of MMES in Ref. 15. It is proposed there that a MMES state in a $n$-qubit system must reach the maximal possible mixedness for each of its subsystem with $[n/2]$–qubit ( $[x]$ takes the integer of $x$ ). Therefore, we can prove a state to be MMES by whether a physical quantity representing the purity averaged over some $[n/2]$–qubit subsystems of the state, defined here as the average subsystem purity of multi-qubit entanglement, reaches its minimum [15]. Specifically, for a $n$-qubit ensemble $S_n=\{1, 2, \ldots n\}$, we can obtain the averaging subsystem purity for a state as

$$\pi_{ME} = \binom{n}{n_A}^{-1} \sum_{|A|=n_A} \pi_A, \tag{1}$$

where $n_A = [n/2]$ and $\pi_A = \text{Tr}_A \rho_A^2$ is the purity of a $n_A$-qubit subsystem $A$ characterized by a $n_A$–bit index $A_1 \ldots A_{n_A}$ with $A_j \in S$ ( $j = 1 \ldots n_A$ ), and the summation runs only over the subsystems with $A_1 < A_2 < \cdots < A_{n_A}$. The marginal density matrix $\rho_A = \text{Tr}_{\bar{A}} |\psi\rangle\langle\psi|$ is the result after the partial trace operation over the complementary subsystem $\bar{A}$ of $A$ in the summation. Here, it obvious that $1/2^{n_A} \leq \pi_A \leq 1$, yet the lower bound $\pi_A = 1/2^{n_A}$ cannot be necessarily satisfied for every $n_A$-qubit subsystem of a certain state, as is not compatible with the criterion that a perfect MMSE must be completely mixed in its arbitrary subsystem in Ref. [15]. In the following, we will only minimize the averaging purity in Eq. (1), without purchasing the completely mixed state in every subsystem. An state satisfying this minimization will be the MMES we can find above the limit.

Merely according to the definition in Eq. (1), the minimization of $\pi_{ME}$ is nearly not available. However, with the local unitary (LU) transformation invariants, a analytically minimizable structure of $\pi_{ME}$ can be obtained. We first introduce five types of LU transformation invariants in Eqs. (2-5) [19] for a state $|\psi\rangle$:

$$T = \langle \psi | \psi \rangle = 1, \tag{2}$$

$$F_i = \sum_{n=x,y,z} \langle \psi | \hat{\sigma}_{in} | \psi \rangle^2, \tag{3}$$

$$F_{ij} = \sum_{n_1=x,y,z} \sum_{n_2=x,y,z} \langle \psi | \hat{\sigma}_{in_1} \hat{\sigma}_{jn_2} | \psi \rangle^2, \tag{4}$$

$$F_{ijk} = \sum_{n_1=x,y,z} \sum_{n_2=x,y,z} \sum_{n_3=x,y,z} \langle \psi | \hat{\sigma}_{in_1} \hat{\sigma}_{jn_2} \hat{\sigma}_{kn_3} | \psi \rangle^2, \tag{5}$$

$$F_{ijkl} = \sum_{n_1=x,y,z} \sum_{n_2=x,y,z} \sum_{n_3=x,y,z} \sum_{n_4=x,y,z} \langle \psi | \hat{\sigma}_{in_1} \hat{\sigma}_{jn_2} \hat{\sigma}_{kn_3} \hat{\sigma}_{ln_4} | \psi \rangle^2, \tag{6}$$

where $\hat{\sigma}_{mn}$ ($m = i, j, k$ and $n = x, y, z$) is a Pauli operator $\sigma_n$ operating on the $m$-th qubit of the state $|\psi\rangle$. It is obviously that such invariants satisfy $F_i \geq 0$, $F_{ij} \geq 0$, $F_{ijk} \geq 0$ and $F_{ijkl} \geq 0$, where the equality acts only when all the expectations terms on the right hand sides of Eqs. (2-6) are zeros.

For a state of $n$-qubit system, we can express it as

$$|\psi\rangle = \sum_{i \leq 2^n} a_i |\varphi_i\rangle, \tag{7}$$

where $a_i$ is the complex coefficient, $|\varphi_i\rangle \in \bigotimes_{j \in S} |\varphi_{ij}\rangle$ and $|\varphi_{ij}\rangle \in \{|0\rangle, |1\rangle\}$. The average subsystem purity $\pi_{ME}$ for this state can be reduced to a structure as

$$\pi_{ME} = C + K(a_1 \ldots a_i \ldots a_{2^n}), \tag{8}$$

where $C$ is a constant and it always has $C > 0$ for arbitrary state; $K$ is the function of the complex coefficients and it can finally be expressed as the combination of the LU transformation invariants in Eq. (2-6). Because of $K \geq 0$, Eq. (8) is a clear minizable structure, in which a state with $K = 0$ and therefore $\pi_{ME} = C$ must be maximally entangled. This criterion will be employed to verify the known or usually used entangled state of 2, 3, 5 and 6 qubits in the follows.

For the case of $n=2$, i.e., the two-qubit system, we can obtain

$$C = 1/2, \quad K = F_1 = \sum_{n=x,y,z} \langle \psi | \hat{\sigma}_{1n} | \psi \rangle^2. \tag{9}$$

In such case, we can easily prove the usually used Bell state $|\varphi_B\rangle_{12} = (|00\rangle + |11\rangle)_{12} / \sqrt{2}$ to be maximally entangled, since $\langle \varphi_B | \hat{\sigma}_{1x} | \varphi_B \rangle = \langle \varphi_B | \hat{\sigma}_{1y} | \varphi_B \rangle = \langle \varphi_B | \hat{\sigma}_{1z} | \varphi_B \rangle = 0$, and then $K = 0$.

For an arbitrary state with $n=3$, we have

$$C = 1/2, \quad \text{and} \quad K = (F_1 + F_2 + F_3)/6. \tag{10}$$

We can prove the experimentally usually used three–qubit GHZ state $|\varphi_G\rangle_{123} = (|000\rangle + |111\rangle)_{123} / \sqrt{2}$ [16] to be a MMSE due to its $K = 0$.

For $n=4$, we have

$$C = 1/3, \quad \text{and} \quad K = K_1 + K_2 \geq 0, \tag{11}$$

where $K_1 = \frac{1}{6}(F_1 + F_2 + F_3 + F_4)$,

and $K_2 = \left[(I_1 + I_2 + I_3 + I_4) - (I_{12} + I_{13} + I_{14} + I_{23} + I_{24} + I_{34})\right]/24$, with $I_i = 1 - F_i$, $I_{ij} = 1 - F_{ij}$. Here, $n_A = 2$ and $C > 1/2^{n_A}$ lead to that $\pi_{ME}$ cannot be minimized to a state in which every state in 2-qubit subsystem is completely mixed, and the conclusion that no 4-qubit MMES exists were obtained from this impossibility [15]. However, the maximally possible entangled state will be obtained under $\pi_{ME} = 1/3$ ($K = 0$), i.e., the relaxation of $\pi_{ME} = 1/2^{n_A}$ to $\pi_{ME} = C$.

According to this criterion, many known 4-qubit highly entangled states can be prove to be MMES. For instance, $|HS\rangle_{1234} = \left[|0011\rangle + |1100\rangle + \omega(|0101\rangle + |1010\rangle) + \omega^2(|0110\rangle + |1001\rangle)\right]/\sqrt{6}$ with $\omega = e^{i2\pi/3}$ and getting greatest possible average entropies, discovered in Ref. [17], has $\pi_{ME} = 1/3$ and therefore it's a MMES. The state employed by Yeo and Chua to implement teleportation and dense coding [20], called genuine four-qubit entanglement state, $|YC\rangle_{1234} = (|0000\rangle - |0011\rangle - |0101\rangle + |0110\rangle + |1001\rangle + |1010\rangle + |1100\rangle + |1111\rangle)_{1234} / 2\sqrt{2}$

is also MMES. It is noted that $\pi_{12}=\pi_{13}=\pi_{14}=1/3$ for $|HS\rangle_{1234}$ and $\pi_{14}=1/2$ for $|YC\rangle_{1234}$, therefore the sate in 2-qubit subsystem of the two MMESs is not necessarily completely mixed. However, the single-qubit reduced states are all completely mixed.

For $n=5$, we have

$$C = 1/4, \quad K = \left(F_1 + \cdots + F_5 + F_{11} + \cdots + F_{45}\right)/40. \tag{12}$$

The state $|\psi\rangle_{12345} = \left(|001\rangle|\phi_-\rangle + |010\rangle|\psi_-\rangle + |100\rangle|\phi_+\rangle + |111\rangle|\psi_+\rangle\right)_{12345}/2$ with $|\psi_\pm\rangle = (|00\rangle \pm |11\rangle)/\sqrt{2}$ and $|\phi_\pm\rangle = (|01\rangle \pm |10\rangle)/\sqrt{2}$, searched in Ref. [17] as a more highly entangled state than 5-qubit GHZ state, has $K = 0$ and therefore is indeed a MMES. At the same time, $C = 1/2^{[5/2]}$ determines that all the subsystems of 1- and 2-qubit of this MMES get the maximal mixedess.

For $n=6$, we can obtain

$$C = 1/8, \quad K = \left(F_1 + \cdots + F_6 + F_{12} + \cdots + F_{56} + F_{123} + \cdots + F_{456}\right)/100. \tag{13}$$

The state searched out in Ref. [18]

$$|\psi_M\rangle_{123456} = \frac{1}{4}[|000000\rangle + |111000\rangle - |001001\rangle + |11000\rangle + |011010\rangle$$

$$- |100010\rangle - |010011\rangle - |101011\rangle + |010100\rangle - |101100\rangle + |011101\rangle$$

$$+ |100101\rangle + |001110\rangle + |110110\rangle + |000111\rangle - |111111\rangle]_{123456}, \tag{14}$$

can be easily to be proven to be MMES due to its $K = 0$.

For the 2, 3, 5 and 6 qubits system described above, $C = 1/2^{n_A}$ is satisfied, and in such case the $n_A$ subsystem of a MMES is completely mixed. However, such lowest bound cannot be reached in the case of 4-qubit, as shown in Fig. 1 and Table. I. This conclusion is consistent with the results in Ref. [17] and gives a explicit lower bound preventing the searching of impossible states.

Now, we search for a MMES for 8-qubit system, which as far as we know has not been discovered. For n=8, we can obtain $C = 6/70$ and

$$K = \frac{11}{280}(F_1 + \cdots + F_8) + \frac{1}{70}(F_{12} + \cdots + F_{78}) + \frac{1}{280}(F_{123} + \cdots + F_{678})$$

$$+\frac{1}{1120}[9(I_1+I_2+\cdots+I_8)+(I_{12}+I_{13}+\cdots+I_{78})$$
$$-(I_{123}+I_{124}+\cdots+I_{678})-(I_{1234}+I_{1235}+\cdots+I_{5678})]. \tag{15}$$

We can measure the entanglement of the experimentally usually used GHZ state and product state by this criterion. For GHZ state, $F_1=\cdots=F_8=0$, $F_{12}=\cdots=F_{78}=1$, $F_{123}=\cdots=F_{678}=0$, $F_{1234}=\cdots=F_{5678}=1$, $K=29/70$ and therefore $\pi_{ME}=1/2$ verify that it is not a MMES; for product state, $F_1=\cdots=F_8=1$, $F_{12}=\cdots=F_{78}=1$, $F_{123}=\cdots=F_{678}=1$, $F_{1234}=\cdots=F_{5678}=1$, $K=64/70$ and therefore $\pi_{ME}=1$ verify that this state is not entangled. Therefore, our criterion is compatible with the common definition of entanglement. In the follows, a 8-qubit MMES will be constructed by LU transformation.

In order to search of a 8-qubit MMES, we can first consider the following quantum state in Eq. (16) in which four 2-qubit subsystems all form 2-qubit Bell states, respectively.

$$|\varphi_B\rangle_{12345678} = \frac{1}{\sqrt{2}}(|00\rangle+|11\rangle)_{12}\frac{1}{\sqrt{2}}(|00\rangle+|11\rangle)_{34}\frac{1}{\sqrt{2}}(|00\rangle+|11\rangle)_{56}\frac{1}{\sqrt{2}}(|00\rangle+|11\rangle)_{78}. \tag{16}$$

It is easy to see that $\pi_{1357}=\mathrm{Tr}_{1357}\rho_{1357}^2=1/16$, i.e., the state in the subsystem composed of qubit (1, 3, 5 and 7) is a completely mixed. Therefore, if $|\psi_M\rangle_{12345678}$ is a MMES, then $|\psi_M\rangle_{12345678}$ and $|\varphi_B\rangle_{12345678}$ are equivalent by a local unitary operation $U_{2468}$ acting on qubits (2, 4, 6 and 8) [19], namely

$$|\psi_M\rangle_{12345678} = U_{2468}|\varphi_B\rangle_{12345678}. \tag{17}$$

By a great deal of mathematical and theoretical calculations, we can find the unitary operation $U_{2468}$ expressed as

$$U_{2468} = \frac{1}{2}\begin{pmatrix}
1 & 0 & 0 & 0 & 0 & 1 & 0 & 0 & 0 & 0 & 0 & -1 & 0 & 0 & 1 & 0 \\
0 & 0 & 1 & 0 & 1 & 0 & 0 & 0 & 0 & -1 & 0 & 0 & 0 & 0 & 0 & 1 \\
0 & 0 & 1 & 0 & -1 & 0 & 0 & 0 & 0 & -1 & 0 & 0 & 0 & 0 & 0 & -1 \\
1 & 0 & 0 & 0 & 0 & -1 & 0 & 0 & 0 & 0 & 0 & -1 & 0 & 0 & -1 & 0 \\
0 & 0 & 0 & 0 & 0 & 0 & 1 & 1 & 0 & 0 & -1 & 0 & -1 & 0 & 0 & 0 \\
0 & 1 & 0 & 1 & 0 & 0 & 0 & 0 & 1 & 0 & 0 & 0 & 0 & -1 & 0 & 0 \\
0 & -1 & 0 & 1 & 0 & 0 & 0 & 0 & 1 & 0 & 0 & 0 & 0 & 1 & 0 & 0 \\
0 & 0 & 0 & 0 & 0 & 0 & 1 & -1 & 0 & 0 & -1 & 0 & 1 & 0 & 0 & 0 \\
0 & 0 & 0 & 0 & 0 & 0 & 1 & 1 & 0 & 0 & 1 & 0 & 1 & 0 & 0 & 0 \\
0 & -1 & 0 & -1 & 0 & 0 & 0 & 0 & 1 & 0 & 0 & 0 & 0 & -1 & 0 & 0 \\
0 & 1 & 0 & -1 & 0 & 0 & 0 & 0 & 1 & 0 & 0 & 0 & 0 & 1 & 0 & 0 \\
0 & 0 & 0 & 0 & 0 & 0 & 1 & -1 & 0 & 0 & 1 & 0 & -1 & 0 & 0 & 0 \\
1 & 0 & 0 & 0 & 0 & -1 & 0 & 0 & 0 & 0 & 0 & 1 & 0 & 0 & 1 & 0 \\
0 & 0 & 1 & 0 & -1 & 0 & 0 & 0 & 0 & 1 & 0 & 0 & 0 & 0 & 0 & 1 \\
0 & 0 & 1 & 0 & 1 & 0 & 0 & 0 & 0 & 1 & 0 & 0 & 0 & 0 & 0 & -1 \\
1 & 0 & 0 & 0 & 0 & 1 & 0 & 0 & 0 & 0 & 0 & 1 & 0 & 0 & -1 & 0
\end{pmatrix}. \quad (18)$$

Then, we obtain the 8-qubit MMES as

$$|\psi_M\rangle_{12783456} = \frac{1}{8}[(|0000\rangle + |0011\rangle - |1101\rangle + |1110\rangle)_{1278}(|0000\rangle + |0111\rangle - |1001\rangle + |1110\rangle)_{3456}$$

$$+ (-|0001\rangle + |0010\rangle + |1100\rangle + |1111\rangle)_{1278}(|0001\rangle + |0110\rangle + |1000\rangle - |1111\rangle)_{3456}$$

$$+ (|0100\rangle - |0111\rangle + |1001\rangle + |1010\rangle)_{1278}(-|0011\rangle + |0100\rangle + |1010\rangle + |1101\rangle)_{3456}$$

$$+ (|0101\rangle + |0110\rangle + |1000\rangle - |1011\rangle)_{1278}(-|0010\rangle + |0101\rangle - |1011\rangle - |1100\rangle)_{3456}]. \quad (19)$$

It can be easily obtained that $K=0$ and therefore this state is verified to be a MMES. The purity of arbitrary $n_A = 4$ qubit reduced subsystem can be obtained as

$$\pi_{1236} = \frac{1}{4}, \ \pi_{1245} = \frac{1}{4}, \ \pi_{1278} = \frac{1}{4},$$

$$\pi_{1347} = \frac{1}{8}, \ \pi_{1358} = \frac{1}{8}, \ \pi_{1468} = \frac{1}{8}, \ \pi_{1567} = \frac{1}{8},$$

$$\pi_{ijkl} = \frac{1}{16}, \ jkl = 1234, 1235, 1237, \cdots, 1678. \quad (20)$$

This equation shows that an arbitrary 4-qubit reduced subsystem of such entangled state is not necessarily completely mixed. Furthermore, it is easy to obtain the purities of other lower-dimension reduced subsystems:

$$\pi_{ijk} = Tr_{ijk}\rho_{ijk}^2 = \frac{1}{8}, \ ijk = 123, 124, \cdots, 678,$$

$$\pi_{ij} = Tr_{ij}\rho_{ij}^2 = \frac{1}{4}, \ ij=12,13,\cdots,78,$$

$$\pi_i = Tr_i\rho_i^2 = \frac{1}{2}, \ i=1,2,3,4,5,6,7,8, \qquad (21)$$

which shows that the 1, 2 and 3 qubits subsystems are all completely mixed. Compared with the Bell state in Eq. (16) and the 8-qubit GHZ state, this state have more nonzero projections in the 256 bases, and is maximally entangled due to $\pi_{ME} = C$, as shown in Fig. (2).

Though the 4 and 8-qubit MMES have been proved and discovered above, the 7-qubit MMES is not found because the constant $C$ in the minimizable expression (7) is hard to extract. We hope that this problem can be solved in the future. However, we find a state of which a reduced 3-qubit sate is not necessarily completely mixed, but a reduced 1- and 2-qubit sate is, as can be shown by the following expressions:

$$\pi_{ijk} = Tr_{ijk}\rho_{ijk}^2 = \frac{1}{8}, \ ijk=123,124,125,126,134,\cdots 467,567; \ ijk \neq 127,457,367,$$

$$\pi_{127} = Tr_{127}\rho_{127}^2 = \frac{1}{4}, \ \pi_{457} = Tr_{457}\rho_{457}^2 = \frac{1}{4}, \ \pi_{367} = Tr_{367}\rho_{367}^2 = \frac{1}{4},$$

$$\pi_{ij} = Tr_{ij}\rho_{ij}^2 = \frac{1}{4}, \ ij=12,13,14,15,16,17,23,\cdots,57,67,$$

$$\pi_i = Tr_i\rho_i^2 = \frac{1}{2}, \ i=1,2,3,4,5,6,7,$$

$$\pi_{ME} = \frac{1}{35}(\pi_{123}+\pi_{124}+\pi_{125}+\pi_{126}+\pi_{127}+\cdots+\pi_{467}+\pi_{567}) = \frac{19}{140}. \qquad (22)$$

On the one hand, this state is entangled more highly than the state having been discovered [18], because it get all the 2-qubit marginal density matrices completely mixed. On the other hand, though $\pi_{ME} = 19/140 > 1/2^{[7/2]}$, not reaching the lowest bound and larger than the numerically discovered result $\pi_{ME} \approx 0.134$ in Ref. [15], the state is completely analytically found without any standard deviation $\sigma_{ME}$, while the result with $\sigma_{ME} = 0$ is $\pi_{ME} \approx 0.136$ in Ref. [15], larger than our result.

Now, synthesizing the case of $n=4$, and the equations (20), (21) and (22), we can conclude that the state in $[n/2]$ subsystem of the 4-, 7- and 8-qubit MMES is not necessarily completely mixed, but all the states in the subsystems of less than $[n/2]$ qubit are completely mixed, as shown in Fig. 1 and Table. I.

The 8-qubit MMES discovered here has many advantages in the applications of teleportation and dense coding. In Eq. (20), we can see that there are many subsystems completely mixed, so it is very convenient to distribute four qubits to Alice and other four to Bob, while such distribution must be elaborate with the usually used GHZ and other states.

In conclusion, we have first introduced a general criterion of maximally multi-qubit entangled states (MMES) by reducing the averaged subsystem purity over all its independent reduced subsystem to be an effective minimizable structure, employing the LU transformation invariants. Such criterion can easily verify the known multi-qubit highly entangled states to be MMES, and in principle it can be applied to arbitrary $n$-qubit entanglement. Also, by the LU transformation we have first discovered a genuine maximally 8-qubit entangled state proved by the criterion. We have revealed that the proved (suspected) MMES of 4, and 8 (7) qubits thought to be nonexistent before, has incompletely mixed subsystems with critical-numbers qubits, but completely mixed subsystem with qubits less than these critical-numbers. We believe this criterion can play important role in determining whether a state discovered in future is maximally entangled, and this genuine 8-qubit MMES can have promising application in the quantum communication and computing.

This work is supported by the National Basic Research Program of China (2012CB921804) and National Natural Science Foundation of China (Grant No. 10974151, 61078002, 61078020, 11104214, 61108017, 11104216, 10902083).

**FIG. 1**. The average subsystem purity $\pi_{ME}$ (rectangle) of a MMES and the lowest bound $1/2^{[n_A]}$ (circle) versus the number of qubit $n$. The lack of $\pi_{ME}$ in $n=7$ is because that a proved 7-qubit MMES has not been discovered in this letter.

**FIG. 2**. The comparison among the 8-qubit Bell state in Eq. (16), GHZ state and the MMES state we discover in this letter. The histograms represent the probabilities for these state to collapse on the 256 bases, and every small rectangle area represents a basis. The three vertical bars are the average subsystem purities $\pi_{ME}$ of the three states, respectively. The dashed dotted line $C$ is the lower bound of $\pi_{ME}$.

**TABLE. I**. The lower bound $C$ of $\pi_{ME}$ and the possible lower bound of a $[n_A/2]$-qubit subsystem, for different $n$. The lack of $C$ for $n=7$ is because that a explicit 7-qubit minimizable structure of $\pi_{ME}$ has not been discovered in this letter.

**Fig. 1**

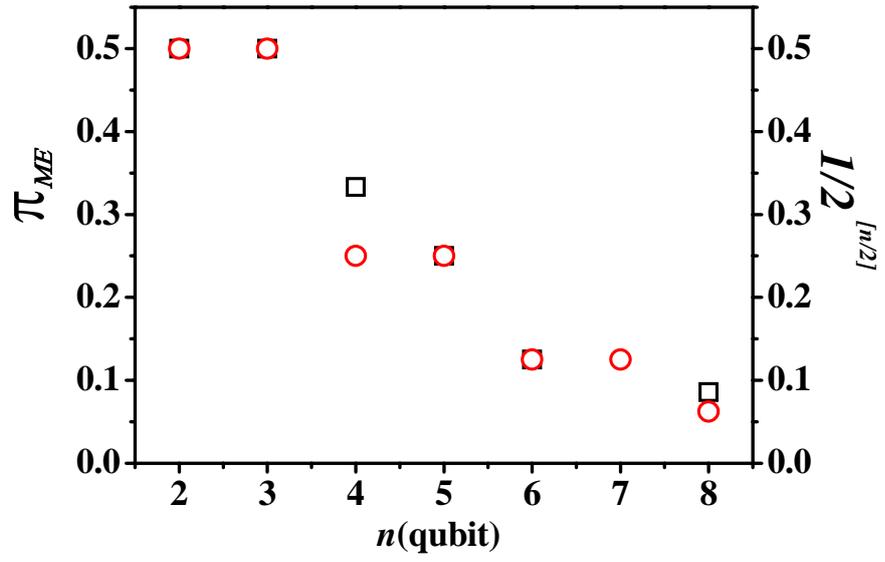

**Fig. 2**

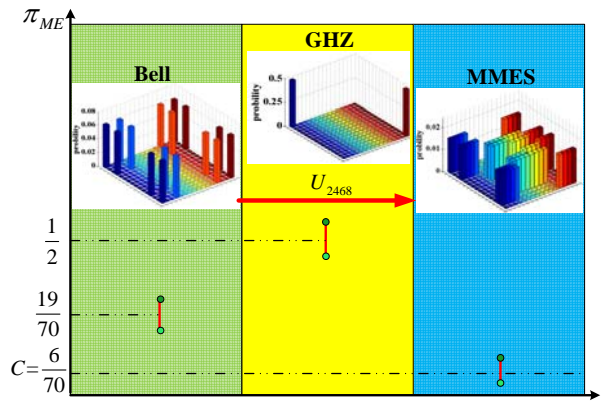

**Table. I**

| n | 2 | 3 | 4 | 5 | 6 | 7 | 8 | ... |
|---|---|---|---|---|---|---|---|---|
| C | 1/2 | 1/2 | 1/3 | 1/4 | 1/8 | ? | 6/70 | ... |
| $1/2^{[n/2]}$ | 1/2 | 1/2 | 1/4 | 1/4 | 1/8 | 1/8 | 1/16 | ... |